\documentstyle{article}
\title{The $\Phi^{4}$ quantum field
in a scale invariant random metric}
\author{ Z. Haba\\Institute of Theoretical Physics, University of Wroclaw,
\\50-204 Wroclaw, Plac Maxa Borna 9, Poland\\e-mail:zhab@ift.uni.wroc.pl}
\date{}
\begin{document}
\maketitle
\begin{abstract}
We discuss $D$-dimensional Euclidean scalar  field interacting
with a scale invariant quantized metric. We assume that the metric
depends on $d$-dimensional coordinates where $d<D$. We show that
the interacting quantum fields have more regular short distance
behaviour than the free fields. A model of a Gaussian metric is
discussed in detail. In particular, in the $\Phi^{4}$ theory in
four dimensions we obtain explicit lower and upper bounds for each
term of the perturbation series. It comes out that there is no
charge renormalization in $\Phi^{4}$ model in four dimensions.
We show that in
a particular range of the scale dimension
there are models in $D=4$ without any divergencies.
\end{abstract}
\section{Introduction}
We discuss $D$-dimensional Euclidean scalar  field interacting
with a quantized scale invariant metric.
 The metric depends on $d$-dimensional
 coordinates.
 The simplest case, which arises from
 a Gaussian metric, will be discussed in detail.
 If $d=2$ then we can give at least
 two non-Gaussian examples of such a scale invariant metric.
 In one example the metric is described
 as a two-dimensional $SL(D,R)$-valued field.
 As the second example we may consider   the
 Polyakov model \cite{pol} in a non-critical
 dimension  which in the conformal gauge
 is reduced to the two-dimensional Liouville model.

In the first model we treat a metric tensor on the Riemannian
manifold (Euclidean formulation)
\begin{equation}
(G)^{AB}=g^{AB}
\end{equation}
as a two-dimensional field $G$ with values in a set of real
symmetric positive definite $D\times D$ matrices $G$. We choose
the metric in a block diagonal form $G^{AB}=\delta^{AB}$ if $A,B>
D-2$ and
 for $A,B\leq D-2$ the tensor $ g^{\mu\nu}({\bf x}_{F})$ is a
$(D-2)\times (D-2)$ matrix depending on  ${\bf x}_{F}\in R^{2}$.
The manifold of positive definite matrices is homeomorphic to  $R\times
SL(D-2,R)/O(D-2)$.
We choose a conformal invariant action for $G$
 \begin{equation}
\begin{array}{l} W(G)=Tr\int d{\bf x}_{F}G^{-1}\partial G
G^{-1}\overline{\partial}G + WZW
\end{array}
\end{equation} where $\partial=\partial_{1}-i\partial_{2}$ is the
holomorphic derivative and $WZW$ denotes the Wess-Zumino-Witten
term \cite{witten}. The conformal invariance of this model has
been shown in refs. \cite{haba}\cite{gawedzki}\cite{berlin}.

The first model is not invariant under the group of
diffeomorphisms of the metric. In the second conformal invariant
model of gravity we consider the string coordinates $X^{\mu}$
interacting with two-dimensional gravity in a way invariant under
general coordinate transformations
\begin{equation}
W=\int d{\bf x}_{F} \sqrt{g}g^{ab}\partial_{a}X^{\mu}\partial_{b}X^{\mu}
\end{equation}
The classical model does not depend on the metric but the
quantum one depends on the conformal Weyl factor. We choose
$g_{ab}=\delta_{ab}\exp \chi $ . A functional integral
over $X$ leads to the effective action
\begin{equation}
W_{eff}(g)=\int d{\bf x}_{F}(\partial_{a}\chi\partial_{a}\chi
+\alpha \exp\chi )
\end{equation}
This Liouville  model is scale invariant
\cite{distler}\cite{dawid}.

In secs.2-3 we discuss general scale invariant models. In sec.4 we
restrict ourselves to the Gaussian metric where we can obtain
detailed estimates on the perturbation series of the
$(\Phi\Phi^{*})^{2}$ interaction ( we discuss the complex scalar
field instead of the real one just for simplicity of the Gaussian
combinatorics). It is shown that quantum fields interacting
with a singular random metric are more regular than the free
fields (a conjecture reviewed in \cite{deser}; see also \cite{salam})

\section{The scalar propagator}
We consider a complex scalar matter field $\Phi$ in $D$
dimensions interacting with gravitons depending only on a
$d$-dimensional submanifold. We split the coordinates as
  $x=({\bf x}_{G},{\bf x}_{F})$ with ${\bf x}_{F}\in R^{d}$.
Without a self-interaction the $\Phi \Phi^{*}$ correlation
function is equal to an average
\begin{equation}
\hbar \int {\cal
D}g\exp\left(-\frac{1}{\hbar}W\left(G\right)\right) {\cal
A}^{-1}(x,y)
\end{equation}
over the gravitational field $g$ of the Green's
function of the operator
\begin{equation}
-{\cal A}=\frac{1}{2}
\sum_{\mu=1,\nu=1}^{D-d}g^{\mu\nu}({\bf x}_{F})\partial_{\mu}\partial_{\nu}+
\frac{1}{2}\sum_{k=D-d+1}^{D}\partial_{k}^{2}
\end{equation}
We repeat some steps of ref.\cite{PLB} (our case here is
simpler and more explicit). We represent the Green's function by means of the proper time method
\begin{equation}
{\cal A}^{-1}(x,y)=\int_{0}^{\infty}d\tau\left(\exp\left(-\tau {\cal A}\right)
\right)(x,y)
\end{equation}
For a calculation of   $\left(\exp\left(-\tau {\cal A}\right)
\right)(x,y)   $ we apply the functional integral
\begin{equation}
\begin{array}{l}
K_{\tau}(x,y)=\left(\exp\left(-\tau {\cal A}\right)
\right)(x,y)=\int {\cal D}x\exp(-\frac{1}{2}\int \frac{d{\bf
x}_{F}}{dt}
  \frac{d{\bf x}_{F}}{dt}-\frac{1}{2}\int g^{\mu\nu}({\bf x}_{F})\frac{dx_{\mu}}{dt}
  \frac{dx_{\nu}}{dt})
 \cr
 \delta\left(x\left(0\right)-x\right)
  \delta\left(x\left(\tau\right)-y\right)
  \end{array}
  \end{equation}
In the functional integral (8) we make a change of variables ($x
\rightarrow b$) determined by Stratonovitch stochastic
differential equations \cite{ikeda}
\begin{equation}
dx^{\Omega}(s)=e_{A}^{\Omega}\left(
x\left(s\right)\right)db^{A}(s)
\end{equation}
where for $\Omega=1,2,....,D-d$
\begin{displaymath}
e^{\mu}_{a}e^{\nu}_{a}=g^{\mu\nu}
\end{displaymath}
and $e^{\Omega}_{A}=\delta^{\Omega}_{A}$ if $\Omega>D-d$.

As a result of the transformation $x\rightarrow b$
the functional integral becomes Gaussian with the covariance
\begin{equation}
E[b_{a}(t)b_{c}(s)]=\delta_{ac}\min(s,t)
\end{equation}
In contradistinction to \cite{PLB}
eq.(9) can be solved explicitly. The solution $q_{\tau}$ of eq.(9) consists of two vectors $({\bf
q}_{G},{\bf q}_{F})$ where
\begin{equation}
{\bf q}_{F}(\tau,{\bf x}_{F})={\bf x}_{F}+ {\bf b}_{F}(\tau)
\end{equation}
and ${\bf q}_{G}$ has the components (for $\mu=1,...,D-d$)
\begin{equation}
q^{\mu}(\tau,{\bf x})=x^{\mu}+\int_{0}^{\tau}
e_{a}^{\mu}\left({\bf q}_{F}\left(s,{\bf
x}_{F}\right)\right)db^{a}(s)
\end{equation}
The kernel is
\begin{equation}
\begin{array}{l}
K_{\tau}(x,y)=E[\delta(y-q_{\tau}(x))]= \cr = E[\delta({\bf y}_{F}
-{\bf x}_{F}-{\bf b}_{F}(\tau))
\prod_{\mu}\delta\left(y_{\mu}-q_{\mu} \left(\tau,x\right)\right)]
\end{array}
\end{equation}
Using eq.(12) and the Fourier representation of the $\delta$-function
we write eq.(13) in the form
\begin{equation}
\begin{array}{l}
K_{\tau}(x,y)=(2\pi)^{-D}\int d{\bf p}_{G}d{\bf p}_{F}
\cr
E[\exp\left( i{\bf p}_{F}\left({\bf y}_{F}-{\bf x}_{F}\right)
+i{\bf p}_{G}\left({\bf y}_{G}-{\bf x}_{G}\right)-i{\bf p}_{F}{\bf
b}_{F}\left(\tau\right)-i\int p_{\mu}e^{\mu}_{a}\left
({\bf q}\left(s,{\bf x}_{F}\right)\right)db^{a}\left(s\right)\right)]
\end{array}
\end{equation}

\section{The scale invariant model}
In general, we cannot calculate the average over the metric
explicitly. However, the scale invariance of the metric is
sufficient for a derivation of the short distance behaviour of
the scalar propagator.

Let us note that     $\sqrt{\tau}b(s/\tau)\simeq \tilde{b}(s)$
where $\tilde{b}$ denotes an equivalent Brownian motion (the
equivalence means that both random variables have the same
correlation functions). Then, using the scale invariance of $e$
with the index $\gamma$  we can write
\begin{equation}
e(\sqrt{\tau}{\bf x}_{F}) \simeq\tau^{-\frac{\gamma}{2}}\tilde{e}({\bf x}_{F})
\end{equation}
Hence, in eq.(12)
\begin{equation}
q^{\mu}(\tau,{\bf
x})=x^{\mu}+\tau^{\frac{1}{2}-\frac{\gamma}{2}}\int_{0}^{1}
\tilde{e}_{a}^{\mu}\left(\tau^{-\frac{1}{2}} {\bf x}_{F}+
\tilde{{\bf b}}_{F}\left(s\right)\right)d\tilde{b}^{a}(s)
\end{equation}
The expectation value over $e$ is
\begin{equation}
\begin{array}{l}
\langle K_{\tau}(x, y)\rangle= \cr
\tau^{-\frac{D-d}{2}(1-\gamma)-\frac{d}{2}} \langle E
\left[\delta\left(\left({\bf y}_{F}-{\bf
x}_{F}\right)\tau^{-\frac{1}{2}}-\tilde{{\bf
b}}_{F}\left(1\right)\right)\delta\left(
\tau^{-\frac{1}{2}+\frac{\gamma}{2}}\left(y-x\right)-
\eta\right)\right]\rangle
\end{array}
\end{equation}
where
\begin{displaymath}
\eta^{\mu}= \int_{0}^{1}
\tilde{e}_{a}^{\mu}\left(\tau^{-\frac{1}{2}}{\bf x}_{F} +
\tilde{{\bf b}}_{F}\left(s\right)\right)d\tilde{b}^{a}(s)
\end{displaymath}

Let $P({\bf u},{\bf v})$ be the joint distribution of $(\eta,\tilde{{\bf
b}}_{F}(1))$
( $P$ does not depend on ${\bf x}_{F}$ because of the translational
invariance). Then, the propagator of the $\Phi$ field is
\begin{equation}
\begin{array}{l}
\hbar\langle {\cal A}^{-1}( x, y)\rangle= \cr \hbar
\int_{0}^{\infty}d\tau\tau^{-(1-\gamma)\frac{D-d}{2}-\frac{d}{2}} P\left(
\left({\bf x}_{G}-{\bf y}_{G}\right)\tau^{(-1+\gamma)/2},\left({\bf
x}_{F}-{\bf y}_{F}\right)\tau^{-\frac{1}{2}}\right)
\end{array}
\end{equation}
Eq.(18) in momentum space has the representation
\begin{equation}
\hbar\langle {\cal A}^{-1}({\bf k}_{G},{\bf k}_{F})\rangle
=\hbar \int_{0}^{\infty}d\tau \tilde{P}
(\tau^{\frac{1-\gamma}{2}}{\bf k}_{G},\sqrt{\tau}{\bf k}_{F})
\end{equation}
where $\tilde{P}$ denotes the Fourier transform of $P$.
Using eq.(14) we may write
\begin{displaymath}
\begin{array}{l}
\hbar\langle {\cal A}^{-1}({\bf k}_{G},{\bf k}_{F})\rangle
\cr
=\hbar\int_{0}^{\infty}d\tau\langle E[\exp i\left(
\sqrt{\tau}{\bf k}_{F}\tilde{{\bf b}}_{F}\left(1\right)
 +\tau^{\frac{1}{2}-\frac{\gamma}{2}}{\bf k}_{G}\eta_{G}\right)]\rangle
 \end{array}
 \end{displaymath}
The dispersion relation (relating the frequency to the wave number)
 is determined  by  (after an analytic continuation $k_{0}\rightarrow
 ik_{0}$)
\begin{displaymath}
(     \langle {\cal A}^{-1}({\bf k}_{G},{\bf k}_{F})\rangle  )^{-1}=0
\end{displaymath}
It can be concluded from eq.(18) that in general the dispersion
relation will be different from the standard one (resulting from
a  wave equation) $k_{0}\sim \vert {\bf k}\vert$. In particular,
we can see that if $\vert {\bf k}_{F}\vert \gg \vert {\bf
k}_{G}\vert $ then $\langle {\cal A}^{-1}({\bf k}_{G},{\bf
k}_{F})\rangle\sim \vert {\bf k}_{F}\vert^{-2}$ whereas if $\vert
{\bf k}_{G}\vert \gg \vert {\bf k}_{F}\vert $     then $ \langle
{\cal A}^{-1}({\bf k}_{G},{\bf k}_{F})\rangle \sim
 \vert {\bf k}_{G}  \vert^{-\frac{2}{1-\gamma}}$.
 In the configuration space, the propagator tends to infinity if both $\vert
{\bf x}_{F}- {\bf y}_{F}\vert$ and $\vert {\bf x}_{G}- {\bf
y}_{G}\vert$  tend to zero. However, the singularity depends in a
rather complicated way on the approach to zero. It becomes simple
if either  $\vert {\bf x}_{F}- {\bf y}_{F}\vert=0$ or $\vert {\bf
x}_{G}- {\bf y}_{G}\vert=0$   . So, if    $\vert {\bf x}_{F}- {\bf
y}_{F}\vert=0$ then we make a change of the time variable
\begin{equation}
\tau=t\vert {\bf x}_{G}-{\bf y}_{G}\vert ^{\frac{2}{1-\gamma}}
\end{equation}
Using eq.(18) we obtain the factor   depending on $\vert {\bf
x}_{G}-{\bf y}_{G}\vert    $ in front of the integral and a
bounded function $A$ of coordinates ,i.e.,
\begin{displaymath}
\langle {\cal A}^{-1}( x, y)\rangle= A\vert {\bf x}_{G}-{\bf
y}_{G}\vert^{-D+2}
\end{displaymath}
If            $\vert {\bf x}_{G}-
{\bf y}_{G}\vert=0$
then we change the time variable
\begin{displaymath}
\tau=t  \vert {\bf x}_{F}-
{\bf y}_{F}\vert^{2}
\end{displaymath}
As a result
\begin{equation}
\langle {\cal A}^{-1}( x, y)\rangle= A \vert {\bf x}_{F}- {\bf
y}_{F}\vert^{-(D-2)(1-\gamma) }
\end{equation}
with a certain bounded function $A$.
We can see that in the ${\bf x}_{G}$ coordinate the singularity
remains unchanged but the propagator is more regular
in the ${\bf x}_{F}$ coordinate.

It is not possible to calculate the probability distribution $P$
exactly. Choosing as a first approximation $\eta\simeq {\bf
b}_{G}(1)$  we obtain \begin{displaymath} P({\bf u},{\bf
v})=(2\pi)^{-\frac{D}{2}}\exp(-\frac{{\bf u}^{2}}{2} -\frac{{\bf
v}^{2}}{2})
\end{displaymath}
In this approximation
\begin{equation}
\hbar\langle {\cal A}^{-1}({\bf k}_{G},{\bf k}_{F})\rangle
=\frac{\hbar}{2} \int_{0}^{\infty}d\tau \exp(- \frac{1}{2}
\tau^{1-\gamma}\vert {\bf k}_{G}\vert^{2}-\frac{1}{2}\tau\vert {\bf k}_{F}\vert^{2})
\end{equation}
In the Higgs model we need the mass term $\int:\Phi\Phi^{*}:$.
The perturbation series for the mass will be finite
if the integral
\begin{displaymath}
\int_{V} d{\bf x}_{F}d{\bf x}_{G}
\vert\langle {\cal A}^{-1}( x, y)\rangle\vert ^{2}
\end{displaymath}
is finite for any bounded region $V$. This integral is
convergent if
\begin{equation}
\int_{\vert k\vert>\epsilon} d{\bf k}_{G}d{\bf k}_{F}\vert\langle { \cal A}^{-1}(k)\rangle\vert^{2}  <\infty
\end{equation}
for any $\epsilon>0$. It can be checked that if $D=4$ then for
any
$\gamma >0$ the integral  (23) is convergent.

In fact, in $D$ dimensions the convergence of the integral (23)
(with the proper time representation
(18)) follows from the convergence of the integral
\begin{equation}
\int_{0}^{1}d\tau_{1}\int_{0}^{1}d\tau_{2}
(\tau_{1}+\tau_{2})^{-\frac{d}{2}}(\tau_{1}^{1-\gamma}
+\tau_{2}^{1-\gamma})^{-\frac{D-d}{2}}
\end{equation}
It is finite if $d+(1-\gamma)(D-d)<4$.
Together with the inequality $\gamma< \frac{1}{2}$
we obtain the inequality $D< 8-d$.

The integral (24) comes directly from  the approximation (22) .
However, we can prove this result in general
under some mild regularity assumptions on the Fourier transform
(19)  of $P$ defined in eq.(18). So,
\begin{displaymath}
\int\vert\langle {\cal A}^{-1}(k)\rangle\vert^{2} dk= \int d\tau
d\tau^{\prime}\tilde{P}(\tau^{\frac{1-\gamma}{2}}{\bf k}_{G},
\sqrt{\tau}{\bf k}_{F})
\tilde{P}(\tau^{\prime\frac{1-\gamma}{2}}{\bf k}_{G},
\sqrt{\tau^{\prime}}{\bf k}_{F}) d{\bf k}_{G}d{\bf k}_{F}
\end{displaymath}
We introduce the spherical coordinates on the plane
$\tau=r\cos\theta$,$\tau^{\prime}=r\sin\theta$. Then, just by
scaling of $r$ we derive  the result (23) under the assumption
that the integral over $\theta$ is finite.

\section{The $(\Phi\Phi^{*})^{2}$ model in a random Gaussian metric}
After the general scale invariant models of gravity
in the previous sections we consider now a Gaussian  model.
For the Gaussian model we can prove explicit upper
and lower bounds on the correlation functions.
We consider a complex scalar matter field $\Phi$ in $D$
dimensions .
Using eq.(12) and the Fourier representation of the $\delta$-function
we write eq.(14) in the form
\begin{displaymath}
\begin{array}{l}
K_{\tau}(x,y)=(2\pi)^{-D+d}\int d{\bf p}_{G}
\exp\left(i{\bf p}_{G}\left({\bf y}_{G}-{\bf x}_{G}\right)\right)
\cr
E[\delta\left( {\bf y}_{F}-{\bf x}_{F}-{\bf b}_{F}\left(\tau\right)\right)
\exp\left(-i\int p_{\mu}e^{\mu}_{a}\left
({\bf q}\left(s,{\bf x}_{F}\right)\right)db^{a}\left(s\right)\right)]
\end{array}
\end{displaymath}
The random variables ${\bf b}_{F}$ and $b^{a}$ are independent.
Hence, using the formula \cite{ikeda}
\begin{displaymath}
E[\exp i\int f_{a}({\bf q}_{F})db^{a}]=
E[\exp(-\frac{1}{2}\int f_{a}f_{a}ds)]
\end{displaymath}
we can rewrite eq.(14) solely in terms of the metric tensor
\begin{equation}
\begin{array}{l}
K_{\tau}(x,y)=(2\pi)^{-D+d}\int d{\bf p}_{G}
\exp\left(i{\bf p}_{G}\left({\bf y}_{G}-{\bf x}_{G}\right)\right)
\cr
E[\delta\left( {\bf y}_{F}-{\bf x}_{F}-{\bf b}_{F}\left(\tau\right)\right)
\exp\left(-\frac{1}{2}\int p_{\mu}p_{\nu}g^{\mu\nu}\left
({\bf q}\left(s,{\bf x}_{F}\right)\right)ds\right)]
\end{array}
\end{equation}
We assume that the metric $g^{\mu\nu}$ is Gaussian with the short
distance correlations
\begin{equation}
\langle g^{\mu\nu}({\bf x}_{F})g^{\sigma\rho}({\bf y}_{F})\rangle
=- D^{\mu\nu;\sigma\rho}({\bf x}_{F}-{\bf y}_{F})=- C^{\mu\nu;\sigma\rho}
 \vert {\bf  y}_{F}-{\bf  x}_{F}\vert^{-4\gamma}
\end{equation}
where $C$ is a scale invariant tensor and $D$ must be positive
definite if the momentum integrals in the final formula are to
exist. Such a requirement contradicts the positive definiteness
of the action for the gravitational field. However, in the
Einstein gravity the conformal modes give a negative contribution
to the action. The model of a conformally flat metric with
$C^{\mu\nu;\sigma\rho}=\delta^{\mu\nu}\delta^{\sigma\rho}$ would
be satisfactory for our purposes ( a proper contour rotation in
the complex space of metrics is needed in order to perform the
functional integral; such an interpretation of the functional
integral over the conformal modes has been considered also in
quantum gravity \cite{hawking}\cite{mazur}). Then, the integral
(17) over $g$ can be calculated
\begin{equation}
\begin{array}{l}
\langle K_{\tau}(x,y)\rangle=(2\pi)^{-D}\int d{\bf p}_{G}
\exp\left(i{\bf p}_{G}\left({\bf y}_{G}-{\bf x}_{G}\right)\right)

\cr
E[\delta\left({\bf y}_{F}-{\bf x}_{F}-{\bf b}_{F}\left(\tau\right)\right)
\exp\left(-\frac{1}{4}\int_{0}^{\tau} p_{\mu}p_{\sigma}
p_{\nu}p_{\rho}D^{\mu\nu;\sigma\rho}\left
({\bf b}_{F}\left(s\right)-{\bf b}_{F}\left(s^{\prime}\right)\right)
dsds^{\prime}\right)]
\end{array}
\end{equation}
The propagator of eq.(3) has the form
\begin{equation}
\langle {\cal A}^{-1}(x,y)\rangle=\int_{0}^{\infty}d\tau
\tau^{-\frac{d}{2}-(D-d)(1-\gamma)/2}F(\tau^ {-\frac{1}{2}}({\bf
y}_{F}-{\bf x}_{F}),\tau^{-\frac{1}{2}+\frac{\gamma}{2}} ({\bf
y}_{G}-{\bf x}_{G}))
\end{equation}
 There is a restriction
on the allowed singularity of the two-point function of the
metric field. So, if the random variable in the exponential
(25) is to be well-defined (without any renormalization)
then the expectation value of its square should be finite
\begin{displaymath}
\langle E[ \left(\int_{0}^{\tau} g^{\mu\nu}\left
({\bf q}\left(s,{\bf x}_{F}\right) \right)ds\right)^{2}]\rangle
<\infty
\end{displaymath}
This expectation value leads to the
integral (if the singularity of $D$ is $\vert {\bf
u}_{F}\vert^{-4\gamma}$    )
\begin{displaymath}
\int d{\bf u}_{F}\int ds\int_{0}^{s}ds^{\prime}(s-s^{\prime})^{-\frac{d}{2}}\exp\Big(-
\frac{1}{2}{\bf u}_{F}^{2}/(s-s^{\prime})\Big)\vert {\bf
u}_{F}\vert^{-4\gamma}
\end{displaymath}
The integral is finite if $\gamma<\frac{1}{2}$.

We calculate the expectation value in the gravitational field (26)
of a product of  any number of propagators. First,  consider the
two-point function of the Wick square which we need for the mass
term in the Higgs model
\begin{equation}
\begin{array}{l}
\langle :\Phi\Phi^{*}:(x) :\Phi\Phi^{*}:(y)\rangle
=\langle\left( {\cal A}^{-1}\left(x,y\right)\right)^{2}\rangle
\cr
=(2\pi)^{-2D+2d}\int d\tau_{1}d\tau_{2}
\int d{\bf p}_{G}d{\bf p}_{G}^{\prime}
\exp\left( i{\bf p}_{G}\left({\bf y}_{G}-{\bf x}_{G}\right)
 +i{\bf p}_{G}^{\prime}\left({\bf y}_{G}-{\bf x}_{G}\right)\right)
\cr
E[\delta\left({\bf y}_{F}-{\bf x}_{F} -
{\bf b}_{F}\left(\tau_{1}\right)
\right)
\delta\left({\bf y}_{F}-{\bf x}_{F}
 -{\bf b}^{\prime}_{F}\left(\tau_{2}\right)\right)
\cr
\exp\Big(-\frac{1}{4}\int_{0}^{\tau_{1}}\int_{0}^{\tau_{1}} p_{\mu}p_{\sigma}
p_{\nu}p_{\rho}D^{\mu\nu;\sigma\rho}\left
({\bf b}_{F}\left(s\right)-{\bf b}_{F}\left(s^{\prime}\right)\right)
dsds^{\prime}
\cr
-\frac{1}{4}\int_{0}^{\tau_{2}}
\int_{0}^{\tau_{2}} p_{\mu}^{\prime}p_{\sigma}^{\prime}
p_{\nu}^{\prime}p_{\rho}^{\prime}D^{\mu\nu;\sigma\rho}\left
({\bf b}_{F}^{\prime}\left(s\right)-
{\bf b}_{F}^{\prime}\left(s^{\prime}\right)\right)
dsds^{\prime}
\cr
-\frac{1}{2}\int_{0}^{\tau_{1}}\int_{0}^{\tau_{2}} p_{\mu}p_{\nu}
p_{\rho}^{\prime}p_{\sigma}^{\prime}D^{\mu\nu;\sigma\rho}\left
({\bf b}_{F}\left(s\right)-{\bf b}_{F}^{\prime}\left(s^{\prime}\right)\right)
dsds^{\prime}\Big)]
\end{array}
\end{equation}
We introduce the spherical coordinates on the
$(\tau_{1},\tau_{2})$-plane $\tau_{1}=r\cos\theta$ and
$\tau_{2}=r\sin\theta$. Next, we rescale the momenta ${\bf
k}_{F}={\bf p}_{F}\sqrt{r}$, ${\bf k}_{F}^{\prime} ={\bf
p}_{F}^{\prime}\sqrt{r}$ , ${\bf k}_{G}={\bf
p}_{G}r^{\frac{1}{2}-\frac{\gamma}{2}}$  and
 ${\bf k}_{G}^{\prime}
={\bf p}_{G}^{\prime}r^{\frac{1}{2}-\frac{\gamma}{2}}$. Then, we
can see that
\begin{equation}
\begin{array}{l}
\langle :\Phi\Phi^{*}:(x) :\Phi\Phi^{*}:(y)\rangle
\cr
=\int d\theta drr r^{-d-(1-\gamma)(D-d)}F(\theta,
r^{-\frac{1}{2}}({\bf x}_{F}-{\bf y}_{F}),
r^{-\frac{1}{2}+\frac{\gamma}{2}} ({\bf x}_{G}-{\bf y}_{G}))
\end{array}
\end{equation}
It follows just by scaling of coordinates (the $r$- integral
scales as twice the $\tau $-integral in eq.(18))
  that for short distances
\begin{equation}
\langle :\Phi\Phi^{*}:(x) :\Phi\Phi^{*}:(y)\rangle
\simeq\left(\langle {\cal A}^{-1}\left(x,y\right)\rangle\right)^{2}
\end{equation}

 We need  to prove that  correlation functions
 (28) are finite and non-zero.
 We show first that the
 bilinear form $(f_{j},\langle{\cal A}^{-1}\rangle f_{l})$
is finite and non-zero on a dense set of functions $f$. For this purpose we choose
\begin{displaymath}
f_{{\bf k}}({\bf x}_{G})=
(2\pi a)^{-\frac{D-d}{2}}\exp(-\frac{a}{2}{\bf x}_{G}^{2}+i{\bf k}{\bf
x}_{G})
\end{displaymath}
Then,
\begin{equation}
\begin{array}{l}
(f_{{\bf k}},\langle{\cal A}^{-1}\rangle f_{{\bf k}^{\prime}})=
(2\pi)^{-D+d}\int_{0}^{\infty}d\tau\tau^{-\frac{d}{2}}\int d{\bf p}_{G}
 \cr E[\delta \left(\tau^{-\frac{1}{2}}\left({\bf y}_{F}-{\bf
x}\right)_{F}-{\bf b}_{F}\left(1\right)\right) \cr
\exp\Big(-\frac{1}{2a}\left({\bf p}_{G}-{\bf k}\right)^{2}
-\frac{1}{2a}\left({\bf p}_{G}-{\bf k}^{\prime}\right)^{2} \cr
-\frac{1}{4}\tau^{2-2\gamma}\int_{0}^{1}p_{\mu}p_{\sigma}
p_{\nu}p_{\rho}D^{\mu\nu;\sigma\rho} \left ({\bf b}_{F}
\left(s\right)-{\bf b}_{F}\left(s^{\prime}\right)\Big)
dsds^{\prime}\right)]
\end{array}
\end{equation}
Both sides depend on ${\bf x}_{F}$ and ${\bf y}_{F}$ because
we integrated out only ${\bf x}_{G}$ and ${\bf  y}_{G}$.
In our estimates  we apply Jensen inequalities in the form (for
real functions $A$ and $f$)
\begin{equation}
E[\exp A]\geq \exp E[A]
\end{equation}
and
\begin{equation}
 E[\exp\Big( -\int_{0}^{1} ds ds^{\prime}f(s,s^{\prime})\Big)]\leq
\int_{0}^{1} dsds^{\prime} E[\exp(-f(s,s^{\prime}))]
 \end{equation}
An upper bound can be obtained by means of the Jensen inequality
(34) expressed in the form
\begin{equation}
\begin{array}{l}
(f_{{\bf k}},\langle{\cal A}^{-1}\rangle f_{{\bf k}^{\prime}})
\cr
 \leq
2\int_{0}^{\infty}d\tau\int_{0}^{1}ds\int_{0}^{s}ds^{\prime}\int
d{\bf u}_{1}d{\bf u}_{2}d{\bf p}_{G}
 \cr
 \tau^{-\frac{d}{2}}\exp\left(-\frac{1}{2a}\left({\bf p}_{G}-{\bf k}\right)^{2}
-\frac{1}{2a}\left({\bf p}_{G}-{\bf k}^{\prime}\right)^{2}\right)
 p(s^{\prime},{\bf
u}_{1})p(s-s^{\prime},{\bf u}_{2}-{\bf u}_{1})
\cr
p\left(1-s,\tau^{-\frac{1}{2}}\left({\bf
y} -{\bf x}\right)-{\bf u}_{2}\right)
\exp\left(-\frac{\tau^{2-2\gamma}}{4}p_{\mu}p_{\sigma}
p_{\nu}p_{\rho}D^{\mu\nu;\sigma\rho}\left ({\bf u}_{1}-{\bf
u}_{2}\right)\right)
\end{array}
\end{equation}
where $p(s,{\bf u})=(2\pi s)^{-\frac{d}{2}}\exp(-{\bf u}^{2}/2s)$.
We can convince ourselves by means of explicit calculations
(using a proper change of variables) that the integral on the
r.h.s. of eq.(35) is finite. For  the lower bound it will be
useful to introduce the Brownian bridge starting from
${\bf x}$ and ending in ${\bf x}+{\bf u}$ \cite{simon} defined on
the time interval $[0,1]$
\begin{equation}
{\bf a}({\bf x},{\bf u},s)={\bf x}+{\bf u}s+{\bf c}(s)
\end{equation}
where ${\bf c}$ is the Gaussian  process starting from $0$ and
ending in  $0$ with the correlation function
\begin{displaymath}
E[c_{j}(s^{\prime})c_{k}(s)]=\delta_{jk}s^{\prime}(1-s)
\end{displaymath}
for $s^{\prime}\leq s$. Then,  the $\delta$  function in eq.(27)
defines the Brownian bridge  and the Jensen inequality (33) takes
the form
\begin{equation}
\begin{array}{l}
(f_{{\bf k}},\langle{\cal A}^{-1}\rangle f_{{\bf k}^{\prime}})
\geq (2\pi)^{-D+d}\int_{0}^{\infty}d\tau\tau^{-\frac{d}{2}}\int
d{\bf p}_{G} \cr
 \exp\Big(-\frac{1}{2a}\left({\bf p}_{G}-{\bf
k}\right)^{2} -\frac{1}{2a}\left({\bf p}_{G}-{\bf
k}^{\prime}\right)^{2}
-\frac{1}{4}\tau^{2-2\gamma}\int_{0}^{1}p_{\mu}p_{\sigma}
p_{\nu}p_{\rho}\cr
E[D^{\mu\nu;\sigma\rho} \left ({\bf
a}\left(0,\tau^{-\frac{1}{2}}{\bf y}_{F}- \tau^{-\frac{1}{2}}{\bf
x }_{F},s\right) -{\bf a}\left(0,\tau^{-\frac{1}{2}}{\bf y}_{F}-
\tau^{-\frac{1}{2}}{\bf x}_{F},s^{\prime}\right)\Big)
]dsds^{\prime}\right)
\end{array}
\end{equation}
where the expectation value in the exponential on the r.h.s.
of eq.(37) is
equal to
\begin{equation}
\begin{array}{l}
\int d{\bf u}\int
ds\int_{0}^{s}ds^{\prime}\omega(s,s^{\prime})^{-\frac{d}{2}}
\exp\Big(- \frac{1}{2}{\bf u}^{2}/\omega(s,s^{\prime})\Big) \cr
\vert
{\bf u}-\tau^{-\frac{1}{2}}s\left({\bf y}_{F}-{\bf x}_{F}\right)+
\tau^{-\frac{1}{2}}s^{\prime}\left({\bf y}_{F}-{\bf x}_{F}\right)
\vert^{-4\gamma}
\end{array}
\end{equation}
where $\omega(s,s^{\prime})=(s-s^{\prime})(1-s+s^{\prime})$.
It is finite if $\gamma<\frac{1}{2}$ (the form (26) of the
graviton two-point function is assumed).

We can make the same estimates for the Wick square
\begin{equation}
\begin{array}{l}
\langle :\Phi\Phi^{*}:(f_{{\bf k}}) :\Phi\Phi^{*}:(f_{{\bf k}^{\prime}})\rangle
\cr
=(2\pi)^{-2D+2d}\int d\tau_{1}d\tau_{2}
\int d{\bf p}_{G}d{\bf p}_{G}^{\prime}
\exp\left(-\frac{1}{2a}\left({\bf p}_{G}+{\bf p}_{G}^{\prime}-{\bf k}\right)^{2}
-\frac{1}{2a}\left({\bf p}_{G}+{\bf p}_{G}^{\prime}-{\bf k}^{\prime}\right)^{2}\right)
\cr
E[\delta\left({\bf y}_{F}-{\bf x}_{F} -
{\bf b}_{F}\left(\tau_{1}\right)
\right)
\delta\left({\bf y}_{F}-{\bf x}_{F}
 -{\bf b}^{\prime}_{F}\left(\tau_{2}\right)\right)
\cr
\exp\Big(-\frac{1}{4}\int_{0}^{\tau_{1}}\int_{0}^{\tau_{1}} p_{\mu}p_{\sigma}
p_{\nu}p_{\rho}D^{\mu\nu;\sigma\rho}\left
({\bf b}_{F}\left(s\right)-{\bf b}_{F}\left(s^{\prime}\right)\right)
dsds^{\prime}
\cr
-\frac{1}{4}\int_{0}^{\tau_{2}}
\int_{0}^{\tau_{2}} p_{\mu}^{\prime}p_{\sigma}^{\prime}
p_{\nu}^{\prime}p_{\rho}^{\prime}D^{\mu\nu;\sigma\rho}\left
({\bf b}_{F}^{\prime}\left(s\right)-
{\bf b}_{F}^{\prime}\left(s^{\prime}\right)\right)
dsds^{\prime}
\cr
-\frac{1}{2}\int_{0}^{\tau_{1}}\int_{0}^{\tau_{2}} p_{\mu}p_{\nu}
p_{\rho}^{\prime}p_{\sigma}^{\prime}D^{\mu\nu;\sigma\rho}\left
({\bf b}_{F}\left(s\right)-{\bf b}_{F}^{\prime}\left(s^{\prime}\right)\right)
dsds^{\prime}\Big)]
\end{array}
\end{equation}
It is clear that we can obtain  lower  and upper bounds
 applying the Jensen inequalities (33) and (34).
 So, for the upper bound
\begin{equation}
\begin{array}{l}
\langle :\Phi\Phi^{*}:(f_{{\bf k}}) :\Phi\Phi^{*}:(f_{{\bf k}^{\prime}})\rangle
\cr
\leq 2(2\pi)^{-2D+2d}\int d\tau_{1}d\tau_{2}
\int_{0}^{1}ds\int_{0}^{s}ds^{\prime}
\int d{\bf p}_{G}d{\bf p}_{G}^{\prime}
\cr
\exp\left(-\frac{1}{2a}\left({\bf p}_{G}+{\bf p}_{G}^{\prime}-{\bf k}\right)^{2}
-\frac{1}{2a}\left({\bf p}_{G}+{\bf p}_{G}^{\prime}-{\bf k}^{\prime}\right)^{2}\right)
\cr
E[\delta\left({\bf y}_{F}-{\bf x}_{F} -
\sqrt{\tau_{1}}{\bf b}_{F}\left(1\right)
\right)
\delta\left({\bf y}_{F}-{\bf x}_{F}
 -\sqrt{\tau_{2}}{\bf b}^{\prime}_{F}\left(1\right)\right)
\cr
\exp\Big(-\frac{1}{4}\tau_{1}^{2} p_{\mu}p_{\sigma}
p_{\nu}p_{\rho}D^{\mu\nu;\sigma\rho}\left
(\sqrt{\tau_{1}}{\bf b}_{F}\left(s\right)
-\sqrt{\tau_{1}}{\bf b}_{F}\left(s^{\prime}\right)\right)
\cr
-\frac{1}{4}\tau_{2}^{2} p_{\mu}^{\prime}p_{\sigma}^{\prime}
p_{\nu}^{\prime}p_{\rho}^{\prime}D^{\mu\nu;\sigma\rho}\left
(\sqrt{\tau_{2}}{\bf b}_{F}^{\prime}\left(s\right)-
\sqrt{\tau_{2}}{\bf b}_{F}^{\prime}\left(s^{\prime}\right)\right)
\cr
-\frac{1}{2}\tau_{1}\tau_{2} p_{\mu}p_{\nu}
p_{\rho}^{\prime}p_{\sigma}^{\prime}D^{\mu\nu;\sigma\rho}\left
(\sqrt{\tau_{1}}{\bf b}_{F}\left(s\right)
-\sqrt{\tau_{2}}{\bf b}_{F}^{\prime}\left(s^{\prime}\right)\right)
\Big)]
\end{array}
\end{equation}
where the r.h.s. of eq.(40) can be expressed by the transition
function for the Brownian motion as in eq.(35).

 For the lower
bound we obtain
\begin{equation}
\begin{array}{l}
\langle :\Phi\Phi^{*}:(f_{{\bf k}}) :\Phi\Phi^{*}:(f_{{\bf k}^{\prime}})\rangle

\geq (2\pi)^{-2D+2d}\int d\tau_{1}d\tau_{2}
\int d{\bf p}_{G}d{\bf p}_{G}^{\prime}
\cr
\tau_{1}^{-\frac{d}{2}}\tau_{2}^{-\frac{d}{2}}
\exp\left(-\frac{1}{2a}\left({\bf p}_{G}+{\bf p}_{G}^{\prime}-{\bf k}\right)^{2}
-\frac{1}{2a}\left({\bf p}_{G}+{\bf p}_{G}^{\prime}-{\bf k}^{\prime}\right)^{2}\right)
\cr
\exp\Big(-E[\frac{1}{4}\int_{0}^{1}\int_{0}^{1} p_{\mu}p_{\sigma}
p_{\nu}p_{\rho}D^{\mu\nu;\sigma\rho}\left
({\bf a}\left(s\right)-{\bf a}\left(s^{\prime}\right)\right)
dsds^{\prime}
\cr
-\frac{1}{4}\int_{0}^{1}
\int_{0}^{1} p_{\mu}^{\prime}p_{\sigma}^{\prime}
p_{\nu}^{\prime}p_{\rho}^{\prime}D^{\mu\nu;\sigma\rho}\left
({\bf a}^{\prime}\left(s\right)-
{\bf a}^{\prime}\left(s^{\prime}\right)\right)
dsds^{\prime}
\cr
-\frac{1}{2}\int_{0}^{1}\int_{0}^{1} p_{\mu}p_{\nu}
p_{\rho}^{\prime}p_{\sigma}^{\prime}D^{\mu\nu;\sigma\rho}\left
({\bf a}\left(s\right)-{\bf a}^{\prime}\left(s^{\prime}\right)\right)
dsds^{\prime}]\Big)
\end{array}
\end{equation}
where we denoted
\begin{displaymath}
{\bf a}(s)={\bf x}_{F}+({\bf y}_{F}-{\bf x}_{F})s+
\sqrt{\tau_{1}}{\bf c}(s)
\end{displaymath}
and
\begin{displaymath}
{\bf a}^{\prime}(s)={\bf x}_{F}+({\bf y}_{F}-{\bf x}_{F})s
+\sqrt{\tau_{2}}{\bf c}^{\prime}(s)
\end{displaymath}
The r.h.s. of eq.(41) can be calculated explicitly using the
correlation function for the Brownian bridge.

 We  compute now higher order correlation functions
\begin{equation}
\begin{array}{l}
\langle \Phi(x)\Phi(x^{\prime})\Phi^{*}(y)\Phi^{*}(y^{\prime})\rangle

\cr

=\langle {\cal A}^{-1}\left(x,y\right) {\cal
A}^{-1}\left(x^{\prime},y^{\prime}\right)\rangle +(x\rightarrow
x^{\prime})
 \cr =(2\pi)^{-2D+2d}\int
d\tau_{1}d\tau_{2} \int d{\bf p}_{G}d{\bf p}_{G}^{\prime}
  \exp\left(i{\bf
p}_{G}\left({\bf y}_{G}-{\bf x}_{G}\right)
 +i{\bf p}_{G}^{\prime}\left({\bf y}_{G}^{\prime}-
 {\bf x}_{G}^{\prime}\right) \right)
 \cr
E[\delta \left({\bf y}_{F}-{\bf x}_{F}-{\bf
b}_{F}\left(\tau_{1}\right)\right) \delta\left({\bf
y}_{F}^{\prime}-{\bf x}_{F}^{\prime} -{\bf
b}_{F}^{\prime}\left(\tau_{2}\right) \right)
\cr \exp\Big( -\frac{1}{4}\int_{0}^{\tau_{1}}\int_{0}^{\tau_{1}}
p_{\mu}p_{\sigma} p_{\nu}p_{\rho}D^{\mu\nu;\sigma\rho}\left ({\bf
b}_{F}\left(s\right)-{\bf b}_{F}\left(s^{\prime}\right)\right)
dsds^{\prime} \cr -\frac{1}{4}\int_{0}^{\tau_{2}}
\int_{0}^{\tau_{2}} p_{\mu}^{\prime}p_{\sigma}^{\prime}
p_{\nu}^{\prime}p_{\rho}^{\prime}D^{\mu\nu;\sigma\rho}\left ({\bf
b}_{F}^{\prime}\left(s\right)- {\bf
b}_{F}^{\prime}\left(s^{\prime}\right)\right) dsds^{\prime} \cr
-\frac{1}{2}\int_{0}^{\tau_{1}}\int_{0}^{\tau_{2}} p_{\mu}p_{\nu}
p_{\rho}^{\prime}p_{\sigma}^{\prime}D^{\mu\nu;\sigma\rho}\left
({\bf x}_{F}-{\bf x}_{F}^{\prime}+{\bf b}_{F}\left(s\right)-{\bf
b}_{F}^{\prime}\left(s^{\prime}\right)\right)
dsds^{\prime}\Big)]     +(x\rightarrow x^{\prime})
\end{array}
\end{equation}
where $(x\rightarrow x^{\prime})$ means the same expression in which
$x$ is exchanged with $x^{\prime}$.
 The four-linear form (42) calculated on   the basis $f$
reads
\begin{equation}
\begin{array}{l}
\langle \Phi(f_{{\bf k}_{1}}) \Phi(f_{{\bf
k}_{3}})\Phi^{*}(f_{{\bf k}_{2}})\Phi^{*}(f_{{\bf k}_{4}})\rangle

\cr
  =(2\pi)^{-2D+2d}\int
d\tau_{1}d\tau_{2} \int d{\bf p}_{G}d{\bf p}_{G}^{\prime}  E[\delta
\left({\bf y}_{F}-{\bf x}_{F}-{\bf b}_{F}\left(\tau_{1}\right)\right)
\delta\left({\bf y}_{F}^{\prime}-{\bf x}_{F}^{\prime} -{\bf
b}_{F}^{\prime}\left(\tau_{2}\right) \right) \cr \exp\left(
-\frac{1}{2a}({\bf p}_{G}-{\bf k}_{1})^{2} -\frac{1}{2a}({\bf p}_{G}-{\bf
k}_{2})^{2} -\frac{1}{2a}({\bf p}_{G}^{\prime}-{\bf k}_{3})^{2}
-\frac{1}{2a}({\bf p}_{G}^{\prime}-{\bf k}_{4})^{2} \right) \cr
\exp\Big( -\frac{1}{4}\int_{0}^{\tau_{1}}\int_{0}^{\tau_{1}}
p_{\mu}p_{\sigma} p_{\nu}p_{\rho}D^{\mu\nu;\sigma\rho}\left ({\bf
b}_{F}\left(s\right)-{\bf b}_{F}\left(s^{\prime}\right)\right)
dsds^{\prime} \cr -\frac{1}{4}\int_{0}^{\tau_{2}}
\int_{0}^{\tau_{2}} p_{\mu}^{\prime}p_{\sigma}^{\prime}
p_{\nu}^{\prime}p_{\rho}^{\prime}D^{\mu\nu;\sigma\rho}\left ({\bf
b}_{F}^{\prime}\left(s\right)- {\bf
b}_{F}^{\prime}\left(s^{\prime}\right)\right) dsds^{\prime} \cr
-\frac{1}{2}\int_{0}^{\tau_{1}}\int_{0}^{\tau_{2}} p_{\mu}p_{\nu}
p_{\rho}^{\prime}p_{\sigma}^{\prime}D^{\mu\nu;\sigma\rho}\left
({\bf x}_{F}-{\bf x}_{F}^{\prime}+{\bf b}_{F}\left(s\right)-{\bf
b}_{F}^{\prime}\left(s^{\prime}\right)\right)
dsds^{\prime}\Big)]+(1,2\rightarrow 3,4)
\end{array}
\end{equation}
where the last term means the same expression with exchanged wave
numbers. We introduce the spherical coordinates on the
$(\tau_{1},\tau_{2})$-plane $\tau_{1}=r\cos\theta$ and
$\tau_{2}=r\sin\theta$. Let us rescale the momenta ${\bf k}={\bf
p}\sqrt{r}$, ${\bf k}^{\prime} ={\bf p}^{\prime}\sqrt{r}$ , ${\bf
k}={\bf p}r^{\frac{1}{2}-\frac{\gamma}{2}}$  and
 ${\bf k}^{\prime}
={\bf p}_{G}^{\prime}r^{\frac{1}{2}-\frac{\gamma}{2}}$. Then, we
can see that the four-point function (42) takes the form
\begin{equation}
\begin{array}{l}
\langle
\Phi(x)\Phi(x^{\prime})\Phi^{*}(y)\Phi^{*}(y^{\prime})\rangle \cr
=\int d\theta drr r^{-d-(1-\gamma)(D-d)}F_{4}(\theta,
r^{-\frac{1}{2}}({\bf x}_{F}-{\bf y}_{F}), r^{-\frac{1}{2}}({\bf
x}_{F}^{\prime}-{\bf y}_{F}^{\prime}), \cr r^{-\frac{1}{2}}({\bf
x}_{F}^{\prime}-{\bf x}_{F}), r^{-\frac{1}{2}}({\bf
x}_{F}^{\prime}-{\bf y}_{F}), r^{-\frac{1}{2}}({\bf
y}_{F}^{\prime}-{\bf x}), \cr
r^{-\frac{1}{2}+\frac{\gamma}{2}} ({\bf x}_{G}-{\bf y}_{G}),
r^{-\frac{1}{2}+\frac{\gamma}{2}} ({\bf x}_{G}^{\prime}-{\bf
y}_{G}^{\prime}), r^{-\frac{1}{2}+\frac{\gamma}{2}} ({\bf
x}_{G}^{\prime}-{\bf y}_{G}),
r^{-\frac{1}{2}+\frac{\gamma}{2}} ({\bf x}_{G}-{\bf
y}_{G}^{\prime}))

\end{array}
\end{equation}
We can conclude from eq.(44) just by scaling that the
singularity of the four-point function is a product
of singularities.

We must define now the $\int (\Phi\Phi^{*})^{2}$ interaction .
First, let us calculate the two-point function of the interaction
Lagrangian
\begin{equation}
\begin{array}{l}
\langle :(\Phi\Phi^{*})^{2}:(x) :(\Phi\Phi^{*})^{2}:(y)\rangle
=\langle\left( {\cal A}^{-1}\left(x,y\right)\right)^{4}\rangle
\cr
=(2\pi)^{-2D+2d}\int E[\prod_{a=1}^{a=4} d\tau_{a}
 d{\bf p}^{a}
\exp\left( i{\bf p}^{a}
\left({\bf y}_{G}-{\bf x}_{G}\right)\right)
\delta\left({\bf y}_{F}-{\bf x}_{F} -
{\bf b}^{a}_{F}\left(\tau_{a}\right)
\right)

\cr
\exp\Big(-\frac{1}{4}\sum_{a,a^{\prime}}
\int_{0}^{\tau_{a}}\int_{0}^{\tau_{a^{\prime}}} p^{a}_{\mu}p^{a}_{\sigma}
p^{a^{\prime}}_{\nu}p^{a^{\prime}}_{\rho}D^{\mu\nu;\sigma\rho}\left
({\bf b}^{a}_{F}\left(s\right)-{\bf b}^{a^{\prime}}_{F}\left(s^{\prime}\right)\right)
dsds^{\prime}\Big)]
\end{array}
\end{equation}
For the lower bound it is sufficient if we let
${\bf x}_{G}={\bf y}_{G}$. Then,

\begin{equation}
\begin{array}{l}
\langle :(\Phi\Phi^{*})^{2}:(x) :(\Phi\Phi^{*})^{2}:(y)\rangle
\cr
\geq(2\pi)^{-2D+2d}\int \prod_{a=1}^{a=4} d\tau_{a}
 d{\bf p}^{a}   \tau_{a}^{-\frac{d}{2}}
\cr
\exp\Big(-\frac{1}{4}E[\sum_{a,a^{\prime}}
\int_{0}^{1}\int_{0}^{1} p^{a}_{\mu}p^{a}_{\sigma}
p^{a^{\prime}}_{\nu}p^{a^{\prime}}_{\rho}D^{\mu\nu;\sigma\rho}\left
({\bf a}^{a}_{F}\left(s\right)-{\bf a}^{a^{\prime}}_{F}\left(s^{\prime}\right)\right)
dsds^{\prime}\Big)]
\end{array}
\end{equation}
where the expectation value in the exponential can be calculated
and the integral on the r.h.s. is finite. For the upper bound of
the interaction $\int_{V} dx(\Phi\Phi^{*})^{2}(x)$ we may take
the test functions of eq.(32) with $2\pi a=V^{-\frac{2}{D-d}}$ and
${\bf k}=0$ in order to approximate the finite volume integral.
With these test-functions we obtain the upper bounds in the same
way as we did it in eqs.(35) and(40)  from the Jensen inequality
(34).

Through an introduction of spherical
coordinates in the $(\tau_{1},...,\tau_{4})$ space  we can show that
for short distances
\begin{equation}
\langle :(\Phi\Phi^{*})^{2}:(x) :(\Phi\Phi^{*})^{2}:(y)\rangle
\simeq\left(\langle {\cal
A}^{-1}\left(x,y\right)\rangle\right)^{4}
\end{equation}
(because the $r$-integral scales as four-times the $\tau$-integral
in eq.(28)).

Let us calculate now the vacuum diagram \begin{equation} I_{2}=
\int_{V}dx\int_{V}dy\langle :(\Phi\Phi^{*})^{2}:(x)
:(\Phi\Phi^{*})^{2}:(y) \rangle
\end{equation}
 corresponding to the second order perturbation expansion in the
 coupling constant. In the momentum space the convergence of this
 diagram follows   from
  the convergence of the integral (for large
 momenta)
  \begin{equation}
 \int dk_{1}dk_{2}dk_{3}\langle {\cal A}^{-1}(k_{1})
{\cal A}^{-1}(k_{2}) {\cal
A}^{-1}(k_{3}){\cal
A}^{-1}(k_{1}+k_{2}+k_{3})\rangle
\end{equation}
On the basis of eq.(47) the integral (49) can be approximated by
 \begin{equation}
 \int dk_{1}dk_{2}dk_{3}\langle {\cal A}^{-1}(k_{1})\rangle
\langle {\cal A}^{-1}(k_{2})\rangle \langle {\cal
A}^{-1}(k_{3})\rangle \langle {\cal
A}^{-1}(k_{1}+k_{2}+k_{3})\rangle
\end{equation}
In the proper time representation the convergence of the integral
(50) depends on the convergence  of the integral (for small
$\tau$)
\begin{equation}
\begin{array}{l}
\int
d\tau_{1}d\tau_{2}d\tau_{3}d\tau_{4}(\tau_{1}+\tau_{4})^{\frac{d}{2}}
\left(-\tau_{1}^{2}\tau_{4}^{2}+\left(\tau_{1}\tau_{3}+\tau_{1}\tau_{4}+\tau_{3}\tau_{4}\right)
\left(\tau_{1}\tau_{2}+\tau_{1}\tau_{4}+\tau_{2}\tau_{4}\right)\right)^{-\frac{d}{2}}
\cr
(\tau_{1}^{1-\gamma}+\tau_{4}^{1-\gamma})^{-\frac{d}{2}+\frac{D}{2}}

\cr\Big(-\tau_{1}^{2-2\gamma}\tau_{4}^{2-2\gamma}
\cr
+(\tau_{1}^{1-\gamma}\tau_{3}^{1-\gamma}
+\tau_{1}^{1-\gamma}\tau_{4}^{1-\gamma}+\tau_{3}^{1-\gamma}\tau_{4}^{1-\gamma}
)
(\tau_{1}^{1-\gamma}\tau_{2}^{1-\gamma}+\tau_{1}^{1-\gamma}\tau_{4}^{1-\gamma}+
\tau_{2}^{1-\gamma}\tau_{4}^{1-\gamma})\Big)^{\frac{d}{2}-\frac{D}{2}}
\end{array}
\end{equation}
We obtained the formula (51) from the representation (19) by
scaling of momenta as in the argument at the end of sec.3 (we
could obtain  lower and upper bounds on the expectation value
(48) using the lower and upper bounds on the Wick powers as in
eqs.(40) and (46); however the explicit integrals are harder for
analysis than eq.(51)).
 In order to investigate the convergence of the integral (51)
it is useful to introduce the spherical coordinates. Then, the
condition for the convergence of the radial
part of the integral (51) reads
\begin{equation}
\frac{3}{2}{d}+\frac{3}{2}(D-d)(1-\gamma)<4
\end{equation}
For $D=4$ this condition has a solution only if $d=1$ and
$\gamma>\frac{4}{9}$ (together with $\gamma <\frac{1}{2}$). If the
condition (52) is satisfied then we obtain quantum field theory
without any divergencies. However, with any $\gamma>0$
 the ultraviolet behaviour is better than
the one with $\gamma=0$. As can be deduced from the dimensional
regularization, if a diagram is only logarithmically divergent
(as for the charge renormalization in $D=4$ model), then after
coupling to quantum gravity it becomes convergent. In $D=4$ with
$\gamma > 0$ (but without the inequality (52)) the vacuum
diagrams are divergent but the ones corresponding to the charge
renormalization are convergent. We calculate, e.g.,  the
four-point function in perturbation expansion at the second order
in the coupling constant

\begin{equation}
\int d^{4}z_{1}d^{4}z_{2}\langle
\Phi(x_{1})\Phi(x_{2})\Phi^{*}(y_{1})\Phi^{*}(y_{2}):(\Phi\Phi^{*})^{2}:(z_{1})
:(\Phi\Phi^{*})^{2}:(z_{2})\rangle
\end{equation}
It is clear from our estimates that the four-point function (53)
will be non-trivial and finite with any $\gamma >0$. We can
calculate  $n$-point functions. It follows that with the lower and
upper bounds established in this section we can obtain finite
lower and upper bounds on each term (eventually after a mass
renormalization) of the perturbation series of the
$(\Phi\Phi^{*})^{2}$ model.

The Gaussian metric with the two-point function (26) may be
unphysical. We applied this model in order to obtain explicit
estimates on correlation functions in the perturbation series. As
shown in sec.3 the regularizing property of the singular random
metric is universal. We expect that with some harder work the
estimates on the perturbation expansion in $(\Phi\Phi^{*})^{2}$ ,
with the two-dimensional gravity of sec.1, are possible as well.

\end{document}